# Computer Aided Detection for Pulmonary Embolism Challenge (CAD-PE)

Germán González, Daniel Jimenez-Carretero, Sara Rodríguez-López, Carlos Cano-Espinosa, Miguel Cazorla, Tanya Agarwal, Vinit Agarwal, Nima Tajbakhsh, Michael B. Gotway, Jianming Liang, Mojtaba Masoudi, Noushin Eftekhari, Mahdi Saadatmand, Hamid-Reza Pourreza, Patricia Fraga-Rivas, Eduardo Fraile, Frank J. Rybicki, Ara Kassarjian, Raúl San José Estépar and Maria J. Ledesma-Carbayo

*Abstract*— *Rationale:* Computer aided detection (CAD) algorithms for Pulmonary Embolism (PE) algorithms have been shown to increase radiologists' sensitivity with a small increase in specificity. However, CAD for PE has not been adopted into clinical practice, likely because of the high number of false positives current CAD software produces. *Objective:* To generate a database of annotated computed tomography pulmonary angiographies, use it to compare the sensitivity and false positive rate of current algorithms and to develop new methods that improve such metrics. *Methods:* 91 Computed tomography pulmonary angiography scans were annotated by at least one radiologist by segmenting all pulmonary emboli visible on the study. 20 annotated CTPAs were open to the public in the form of a medical image analysis challenge. 20 more were kept for evaluation purposes. 51 were made available post-challenge. 8 submissions, 6 of them novel, were evaluated on the 20 evaluation CTPAs. Performance was measured as per embolus sensitivity vs. false positives per scan curve. *Results:* The best algorithms achieved a per-embolus sensitivity of 75% at 2 false positives per scan (fps) or of 70% at 1 fps, outperforming the state of the art. Deep learning approaches outperformed traditional machine learning ones, and their performance improved with the number of training cases. *Significance:* Through this work and challenge we have improved the state-of-the art of computer aided detection algorithms for pulmonary embolism. An open database and an evaluation benchmark for such algorithms have been generated, easing the development of further improvements. Implications on clinical practice will need further research.

## I. Introduction

Pulmonary embolism is formed when a portion of a blood clot breaks off from the wall of a vein and travels through the blood stream, passes through the right side of the heart (right atrium and right ventricle), and becomes lodged in a pulmonary artery, causing a partial or complete obstruction. Such an obstruction impedes blood flow to the affected portion of the lung resulting in dead space. As a result, the affected portion of the lung does not participate in its primary function of oxygenating the blood and removing carbon dioxide from the de-oxygenated blood. Poor oxygenation results in poor delivery of oxygen to the vital organs, which may subsequently malfunction or fail to function. The increase pulmonary resistance can evolved in to right heart failure in some subjects if treatment is delayed or inadequate resulting in increased morbidity and mortality [1]. Therefore, a rapid diagnosis is of extreme importance. Pulmonary emboli affect between 300,000-600,000 Americans, resulting in 12,000-80,000 deaths/year [2].

Computed tomography pulmonary angiography (CTPA) is accepted as the diagnostic imaging study of choice to confirm the clinical suspicion of acute PE [3],[4]. The diagnosis of PE by CTPA is made by the identification of a filling defect in a pulmonary artery. Usage studies suggests that by using CAD as a second opinion, radiologists can improve their sensitivity with a minimal decrease in specificity [5]–[8]. In a retrospective study [9], out of 6769 consecutive CTPA scans, 703 studies were positive for PE by a panel of experts, and 44 studies were not originally reported as PE-positive. A CAD algorithm found at least one embolus in 77.4% of such 44 studies. 14 patients with missed PE who were not receiving anticoagulation therapy developed further episodes of pulmonary embolism.

Despite the potential advantages of using CAD as a second reader of CTPA exams, few if any, radiologists use such systems. The greatest drawback of CAD is the high false positive rate, often in the range of 4 false positives per scan [10]. Such false positives are often located in pulmonary veins or airspace consolidations [6]. The relatively recent advent of artificial intelligence and convolutional neural networks might improve such performance metrics, and thus the renewed interest in CAD for PE.

Among the bottlenecks of the development of such algorithms is the access to data. During the last decade the medical imaging community has been proposing "challenges": making a public database of medical images with their associated reference standard and defining an evaluation metric in order to help objective algorithm comparison. Along those lines, we proposed the CAD-PE challenge [11] alongside with the IEEE International Symposium of Biomedical Imaging (ISBI) with the goals of a) generating a public database of CTPAs on which to test CAD-PE algorithms and b) develop new algorithms that lower false positive rates of current CAD-PE solutions while keeping the same sensitivity. The challenge remained open after the Symposium to enable a higher number of participants.

In this paper we describe the state of the art on CAD for PE, the CAD-PE challenge dataset, the proposed evaluation metric used to compare CAP-PE algorithms and we finally conduct a comparative studies between eight different CAD-PE algorithms on such database, as a result of the challenge and the on-going testing period.

## II. RELATED WORK

The CAD community has been developing solutions for the automatic detection of PE in CTPA images for the last decade [12]. The metric that has been used to perform the evaluations is often the per-embolus sensitivity and the number of false positives per scan. Current sensitivities range from 61% to 90% at an approximate rate of 4.5 false positives per scan, as shown in table S1. Most follow a three-step method. First the algorithm find potential image locations were emboli can be present. Second, such locations are described as a set of features, or numbers, that can be used to discern between emboli and confounding structures. Finally, such distinction is made with a machine-learning technique.

Most methods extract candidates by segmenting the pulmonary vessels and consider each point within the vessel as a candidate of PE. Vessel segmentation has been performed by a) threshold and vessels tracking [13], [14], b) expectation-maximization analysis [15], [16], c) hysteresis thresholding [17] or d) multiple active contours [18]. Some methods [19], [20] segment arteries from veins using the structures of the airway tree as cues. An alternative to vessel extraction techniques, is the direct location of candidate points from the properties of the emboli directly, with the "tobogganing" technique [21]–[23].

In the second step most algorithms compute a set of features from a region of interest (ROI) centered at the candidate point. Such features are often based on local contrast, grey level intensity, shape, boundaries, size and gradient change. Further features are Gabor filters to detect texture features in the image [18] and the incorporation of geodesic distance maps to establish associations between PE candidates [24]. The large number of extracted features usually offers redundant information. Some systems apply feature selection methods to identify the most useful features, such as stepwise regression [13], [16] or genetic algorithms [22].

Finally, a classifier is designed to distinguish between candidates within emboli and candidates in other structures using the selected set of features as input. Several standard machine learning algorithms have been proposed, such as support vector machines (SVMs) [24], decisions trees [13], [17], Linear discriminant analysis (LDA) [16], [25], decision rules [15] or artificial neural network (ANN) [18], [23]. However, almost all these methods do not properly handle the particular properties of PE detection: an imbalanced training dataset (there is often a ratio of 1 positive training sample to hundreds of negative ones) and the fact that several positive training points can belong to the same embolus. To address such issues, the work of Park et al. [22] combines together ANN and a multi-feature based k-nearest neighbor classifier and the work of Bi et al. [21], [25] uses bagging classifications approaches such as Multiple Instances Learning (MIL).

In contrast to the paradigm of feature design, selection and classification, the work of [23], [26], [27] compute a multi-planar representation of the region of the embolus and uses a convolutional neural network to both learn the features and the discrimination function.

In this paper, the authors from University of Alicante presented an alternative method to the three-stage algorithms. Their convolutional neural network analyzes the whole image to find the emboli, without a prior step of candidate selection.

In table S2 we review clinical studies performed with CAD for PE methods. Their performance depend strongly on the characteristics of the emboli, such as size distributions, percentage of occlusion of an artery by PE, and the diameter of the artery being occluded [20]. Image quality is another source of noise [6], [28], with the degree of contrast filling and motion artifacts being the main sources of noise [29]. The presence of other pulmonary diseases further complicates the detection of PE [15]. The work of [30] further investigates how CAD solutions behave with iterative reconstruction techniques, showing a decrease of sensitivity and false positives per scan when using iterative reconstruction.

Several CAD systems show high sensitivity [17], [31]. However, they are often evaluated in data sets with a small number of emboli. Another study did not include data sets with motion artifacts, pulmonary diseases or unsatisfactory vascular opacification [28], biasing the evaluation of its clinical performance, as reported by [10]. Systems that employed large dataset with realistic information [13], [15], [32] often report high rates of false positives at high sensitivity values.

### A. Objectives and contributions

This work has a multiple objective: 1) to describe a public dataset of 91 CTPA annotated scans to ease the development of CAD for PE algorithms; 2) to define an objective benchmark to compare the already existing algorithms; and 3) to present and compare current solutions for CAD for PE in the context of a challenge where training data was provided to all participants and testing was conducted in a blinded and centralize fashion and 4) develop new algorithms for CAD for PE. It has been proven that CAD for PE performance depends on the testing data set [12]. Therefore, the existence of an open annotated reference dataset for PE provides the opportunity to make a direct comparison among the performance of different systems. Such comparison is automatic, not requiring expert input, making it objective. Both the data and the evaluation metric are open to the community in the form of a challenge, which was held as part of the IEEE International Symposium on Biomedical Imaging [11] and continued to be available afterwards. Three participants submitted results to the challenge and five other participants submitted afterwards.

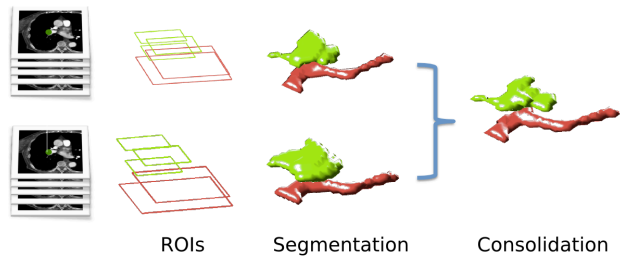

Figure 1 Reference standard generation. Each radiologist marks a set of ROIs in each CTPA. The ROIs are turned into a segmentation of the emboli via a semi-supervised method. The segmentations are adjudicated into the final reference standard via the method of STAPLE.

### III. MATERIALS AND METHODS

#### A. Dataset and Reference Standard

91 computed tomography pulmonary angiography studies positive for PE were collected retrospectively from the clinical data acquired at six different hospitals associated with Unidad Central de Radiodiagnóstico in Madrid, Spain, a central system that concentrates the radiology services of the Madrid region. Protected health information (PHI) was removed from the studies. All scans were obtained in a caudocranial direction from the level of the diaphragm to the lungs apices with a single breath hold. All studies were performed with SIEMENS Somaton Sensation 40 scanners. The institutional CTPA protocol was followed at each site. Image pixel size ranges from 0.58 to 0.85 mm and reconstruction slice thickness between 0.75 and 1.5 mm/slice. We did not exclude any CTPA study because of the presence of other pulmonary diseases.

A board-certified radiologist with more than 10 years of experience diagnosed each CTPA study as PE-positive or negative. The first 40 PE-positive studies from distinct subjects (17 Female, age 65 ± 18 years) were selected to generate the CAD-PE challenge dataset. All selected studies were deemed of adequate clinical quality. The noise as measured in the descending aorta at the level of the bifurcation of the pulmonary artery was 45.94 ±16.1 HU. The reference standard was constructed by three experienced radiologists: A.K. a board certified radiologist with over 15 years of experience reading CTPAs; P.F. head of the radiology unit of one of the hospitals, with more than 20 years of clinical experience and S.H., thoracic radiologist with more than 19 years of clinical practice, the last 10 specialized on thoracic radiology. The scans were reviewed using the Osirix software using a DELL U2410 or a LG 235V monitor, meeting the requirements for reading clinical scans.

Each radiologist independently detected all emboli visible in an image by generating a region of interest (ROI) in each axial slice of each visible embolus. Sagittal and coronal images were used to confirm the presence and absence of emboli. From such markings, a semi-automated method for embolus segmentation was instantiated. Such segmentation consists of a thresholding step based on Hounsfield units (HU), followed by a closing operation and connected component analysis. An image analysis expert removed spurious pixels and delineated emboli borders on each segmentation. A final reference segmentation was established by consolidating the three segmentations with the method of STAPLE [33]. Please note that the method of STAPLE applied to the per-radiologist segmentation resolves discrepancies among different radiologists. The consolidated segmentation was manually inspected to detect and remove spurious voxels. This process is shown in Figure 1.

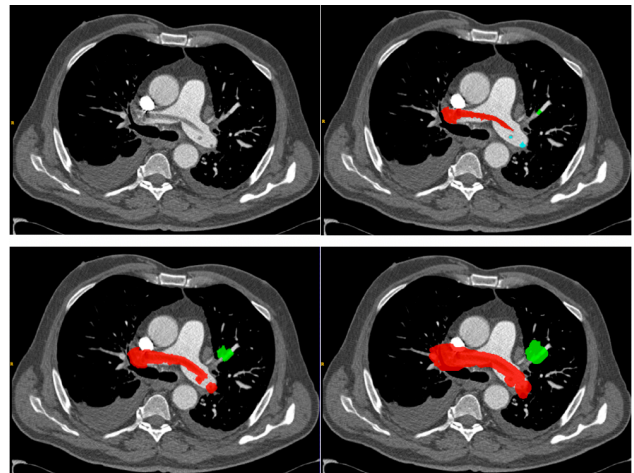

Figure 2 Example of reference standard. Case #1 of the training set. Top-left: axial slice displaying a saddle embolus. Top-right: reference standard with an ε=0mm. Bottom-left ε=2mm. Please note that the red and blue emboli of the top-right image have merged. Bottom-right ε=5mm.

Each connected component within the consolidated segmentations was considered an individual embolus. The resulting dataset had 235 emboli (105 in the first 20 cases, 130 on the evaluation dataset). Average embolus size was $3,43 \cdot 10^3 \pm 8,6 \cdot 10^3$ mm$^3$ (minimum size 8.63 mm$^3$, maximum size $6,34 \cdot 10^4$ mm$^3$). The minimum number of emboli per case was 0 and the maximum 21.

The reference standard was further modified by adding two tolerance margins, correspondent to the epsilon values, ε = 2 mm and ε = 5 mm from the border of each embolus. Such modification is rationalized by the fact that if the detection lies close to an embolus, it can still be useful when using CAD as a second reader. As some emboli merged, connected component analysis was re-run to re-compute the number of emboli. For ε = 2 mm there were 201 emboli (87 in the first 20 cases, 114 on the evaluation dataset) with an average size of $1,02 \cdot 10^4 \pm 2,24 \cdot 10^4$ mm$^3$ (minimum size 251 mm$^3$, maximum size $1,63 \cdot 10^5$ mm$^3$). For ε = 5 mm there were 155 emboli (65 in the first 20 cases, 90 on the evaluation dataset) with an average size of $2,9 \cdot 10^4 \pm 5,17 \cdot 10^4$ mm$^3$ (minimum size $1,45 \cdot 10^3$ mm$^3$, maximum size $3,24 \cdot 10^5$ mm$^3$). Examples of the reference standard with the tolerance margins are shown in Figure 2.

The additional 51 CTPA studies were analyzed similarly, but by only one board-certified radiologist. Only the ε = 0 mm is provided for such studies. The research was carried out following the principles of the Declaration of Helsinki.

*B. Evaluation Metric*

Each algorithm was evaluated by finding its sensitivity, average number of false positives per scan, the positive predictive value and the average and the standard deviation distance from false detections to their closest emboli. We used free receiver-operator characteristic curves to assess the trade-off between sensitivity and false positives per scan for different methods.

Sensitivity was defined as the number of emboli detected divided by the number of total emboli in the test scans. The average number of false positives per scan was computed by adding all false positive detections and dividing by the number of scans in the testing set. Positive predictive value was measured as the ratio of positive detections with respect to all detections in all scans.

True positives were defined as any location provided by the algorithm that was within the reference standard emboli. In the case that several detections fell within the same embolus, only the detection with the highest confidence score, as provided by each method, was taken into account. False positives were defined as locations outside the emboli. False negatives were defined as emboli without any detection inside them.

*C. CAD-PE Challenge Rules*

Registration was open to academic institutions, companies and individuals. After registering on the Challenge website[11], teams could download the complete dataset. The CTPAs were separated in two batches: 20 scans for training and another 20 scans for testing. A reference standard was provided for the training batch. The organizers kept a similar reference standard for the testing set of CTPAs.

Each team was requested to provide the detection of PEs in the testing batch in a text file for evaluation, and an abstract with the description of their algorithm. The text file submitted was standardized: each line had to present the information of each detected embolus, detailing the number of scan, the embolus location expressed in pixels and a confidence score associated. The participants also had to provide a threshold on that confidence score, which is used to compute the evaluation metrics.

Participants' algorithms were expected to automatically mark the position of the emboli by placing markers within them. No user interaction was allowed. The evaluation software for the training set was provided in Matlab. We recommended that the participants do leave-one-out cross-validation for training.

*D. Challenge Participants*

65 teams registered and downloaded the data. A total of three teams submitted their results in time for the challenge day on April 7th 2013 and five other participants afterwards. In this section we briefly describe the methods of the participants categorized by the training dataset used for their development. Five methods used the 20 training cases exclusively. One method was a commercial solution. Two other methods used more than 20 cases for their training.

*1) Methods trained on the challenge database*
*Naïve approach (SVM-Features)*
This method was developed by the challenge organizers to serve as reference approach. This approach follows the standard methodology described in Section 2: vessel segmentation, feature extraction and feature classification. Vessel segmentation was performed by segmenting the lungs using the Chest Imaging Platform [34], [35]. The images of the segmented lungs were intensity equalized to normalize them with respect to image acquisition parameters. Candidate points were generated using the particle-based method of [36]. The candidate points were placed along the arterial and venous lung vasculature and had a vector indicating the direction of the vessel. For each candidate point SVM-Features's method generated a set of features that included 3-D and 2-D based measurements computed by resampling the images along and across the vessel of interest. Example features were: intensity, gradient magnitude, hessian eigen-values, curvature and laplacian. The features were then used to train a support vector machine (SVM) classifier with a radial basis function as kernel on 300 positive candidate points (within PE emboli) and 300 negative candidate points chosen at random per scan. Leave one subject out cross-validation was used to optimize the parameters of the SVM.

*LMNIIT*
First, lungs were segmented using a lung HU value threshold and the K-means clustering algorithm. Vessels were subsequently segmented using expectation maximization and morphological operations. Over the segmented region they computed shape features: eigenvalues of Hessian Matrix, texture features: gray-level entropy matrix (GLEM) and gray-level co-occurrence matrix (GLCM) and basic intensity distribution features: kurtosis and skewness. The Sequential Floating Forward Selection method (SFFS) was employed to determine the best features among the total set and reduce the problem dimension. The imbalanced data set was handled by oversampling the positive training set, i.e. adding new samples created by rotation of the true PEs through several angles. Support vector machine (SVM) with a linear kernel was used for the classification of the segmented instances.

*Universidad Politecnica de Madrid (UPM)*
The departure point of the UPM followed the SVM-Features submission. First, the lungs were segmented including the central vasculature. The vessels were segmented using the scale-space particle method [36]. For each particle, 3D histograms were extracted. To deal with the imbalanced data set of candidates and the fact that several particles could belong to the same embolus UPM's method used multiple-instance learning combined with AdaBoost, using random forests as weak classifiers. The main idea behind multiple instance learning is that the classifier receives a set of bags, each of them composed of close-by particles. Each bag is labeled as positive or negative depending on its particles' labels. In this this method did not classify individually particles but sections of the image.

This technique resulted in a large reduction of the false positive rate.

*FUM-MvLab*
The proposed algorithm contains two parts: emboli modeling and probability generation based on the error measurement.

Emboli modeling: All of emboli regions indicated in ground truth are used as training samples. For every pixel in these regions, a 22 dimension feature vector is extracted. These features are obtained from 2-D and 3-D structure information, such as CT Hounsfield value, statistics, statistical moments, pixel value fluctuation, bottom hat transform, eigenvalues of the Hessian matrix in different scales, circularity, vesselness, degree of curvature, and size of clots. PCA is used to identify the dominant components. Since the proposed transformation by PCA is linear, the suggested model is convex and inherently has generalization. Generally the eigenvectors define the main axes of distribution of training samples, while, eigenvalues are a measure of the dispersion of training samples along with the corresponding eigenvectors. A method for modelling of distribution of training samples is the elimination of weak eigenvectors. If a feature vector belongs to the foreground, the corresponding coefficients of weak eigenvectors will be small while once it belongs to the background, that would not happen. Therefore, we can use the elimination of weak eigenvectors as an effective strategy to distinguish the foreground and background pixels (for more details, see [37]).

Lung segmentation is performed by adaptive thresholding, morphological operations (for including small vessels), and 3D region growing to include the main arterial tree in the mediastinum. After lung segmentation in the test phase, the, feature vectors of all ROI pixels are obtained. Then, a PCA-based function error is computed for each pixel. All error values are stored in the corresponding position of the error image φ. Obviously, the background pixels of φ, have higher error values compared to the foreground ones. Theretofore, we can separate foreground from background by using an adaptive threshold value. Since in the challenge PE-CAD, choosing only one pixel from each clot was enough (because emboli segmentation was not the main goal); in the proposed algorithm, we used connected component analysis with adaptive thresholding to select only one candidate for each clot (we also tested our method for emboli segmentation by using the FUMPE dataset [38]).

*Universidad de Alicante – 2D*
This method uses a simplified version of the U-Net segmentation network [39] on axial slices to segment the emboli directly. The network is trained in a two-step manner. First, we optimize training meta-parameters (learning rate, decay rate, number of epochs and batch size) using 15 scans as training and 5 for validation. Then, we use the 20 scans to train the model with the already set meta parameters. Input images are clipped between the values [-200,500] HU and normalized in the range [0,1]. The loss function is the binary cross-entropy between the network output and the reference standard. The optimization method is Adam, with a learning rate of 0.0005. We turn the output probabilities into emboli coordinates with the following method: 1) Threshold at a value of 0.5 2) Binary closing with a kernel of 5x3x3. 3) The unconnected segmentation clusters are labeled to get the different emboli inside the scan. 4) For each label (emboli) we obtain the coordinate closest to the center of mass from all coordinates of the clot that are equidistant from the edges. This last method results in a list of coordinates, one for each predicted emboli.

*2) Commercial methods*
*Mevis*
Mevis used their previously developed automated CAD for PE algorithm for this challenge. This algorithm finds filling defects in pulmonary arteries greater than 4mm in diameter, and automatically computes vessel diameter, the percentage of occlusion and the average density of the opacified blood in the artery. The algorithm follows the structure of other CAD-PE solutions in that it performs lung segmentation, followed by vessel segmentation, location of low attenuation areas within the vessels and a final PE segmentation. The algorithm does not mark defects in pleural effusions or defects within the mediastinum. False positives can be found in filling defects in the veins and in soft-tissue densities or lymph nodes outside the vessels.

*3) Methods trained in extended databases*
*ASU-Mayo*
The suggested system was based on a unique PE representation coupled with CNNs, consisting of 4 stages: First, the lungs were segmented from the CTPA dataset using a method based on intensity thresholding and morphological operations. Second, a set of PE candidates was generated in the segmented lungs using the tobogganing algorithm. Third, a novel vessel-aligned multi-planar image representation was utilized to capture image information from each PE candidate. The suggested image representation offered three advantages: (1) efficiency and compactness---concisely summarizing the 3D contextual information around an embolus in only 2D image channels, (2) consistency---automatically aligning the embolus in the 2-channel images according to the orientation of the affected vessel, and (3) expandability---naturally supporting data augmentation for training convolutional neural networks (CNNs). Fourth, the resulting 2-channel images were fed to a CNN for feature extraction and candidate classification. The suggested system contrasted with existing systems, wherein a traditional hand-crafted feature design is used for characterizing PEs. A more complete description of the algorithm can be found in [40], [41].

*Universidad de Alicante – 2.5D*
This method follows the experimental setup of the Universidad de Alicante – 2D method, but with the following changes: 1) The input of the segmentation network is not a single axial slice, but five of them (two above the plane that we are segmenting and two bellow such plane). The output of the network is the segmentation of the central plane. 2) We use a total of 60 scans for training, 20 from the training set of the challenge and 40 from the CAD-PE dataset that were not available during the challenge phase. The training set is split into 55 scans for training and 5 for validation. The model performing best on the validation set is used to generate the output coordinates with the same reduction methods as our 2D method.

*4) Methods that self-evaluate on the CAD-PE database*

The work of Lin et al. [42] propose a method for the detection of PE on CTPA images using three convolutional neural networks: a candidate proposal subnet, a 3D spatial transformation subnet and a false positive removal subnet. Their method achieves a sensitivity of 78.9%, 80.7% and 80.7% at 2 FP/scan at 0mm, 2mm and 5mm localization error. While their results were excellent, the measurements were performed with respect to their own segmentation masks, generated by a radiologist with 10 years of experience. A fair comparison with the rest of the methods requires the masks of the challenge to be used.

## IV. RESULTS

Figure 3 shows the sensitivity of the different methods with respect to the number of false positives per scan for the labels with an epsilon of 5mm. The three best performing algorithms were the two submissions of the Universidad de Alicante, followed by the commercial solution of Mevis and the team of ASU-Mayo. The operating point of the methods might not be optimal, since it was selected on the training dataset.

In the supplementary material we evaluate the sensitivity, false positives per scan, positive predictive value and distance from false positives to emboli for the participations, evaluated on the 20 test cases of the dataset (see Table S3). Further, we evaluate the performance of the methods per individual CT scan, where we show disparity between methods – some generate more false positives at given scans (see Table S4. Also, in the supplementary material, we present an evaluation of the performance of the methods with respect to the epsilon parameter used to obtain the reference standard (see Tables S3, S5, and figures 3, S1 and S2).

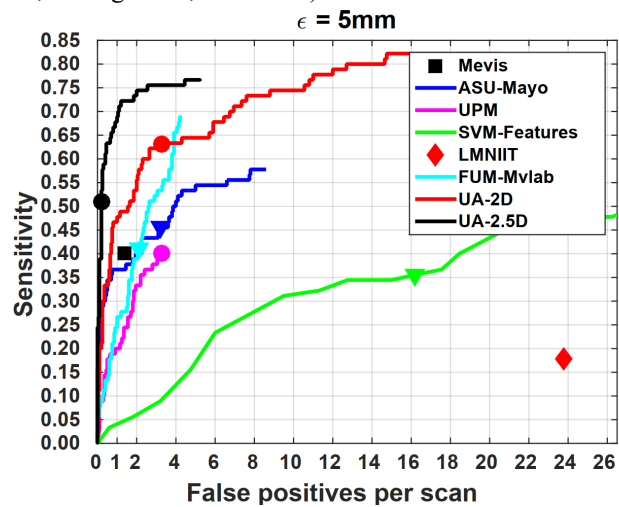

Figure 3 – Evaluation of the different methods on the 20 evaluation scans. The participants were blind to the reference standard. Y-axis: per-embolus sensitivity. X-axis – number of false positives per scan. Evaluation performed using the 5mm epsilon. The dots in the curves represent the operating point selected by the submission.

## V. DISCUSSION

Research on computer aided detection algorithms is often carried out using databases that are closed to the laboratory doing the research, biasing the development and objective evaluation of the algorithms. Challenge databases and evaluation metrics have appeared in the last decade, enabling more objective comparison of algorithms' results. In this paper we have presented the first open database of annotated CTPAs and a concise evaluation metric that can be used to further research in the development of pulmonary embolism computer aided detection methods. Eight different algorithms have been evaluated with the database following the proposed evaluation methodology. The results show good performance of CNN-based CAD solutions, achieving a per-embolus sensitivity of 0.75 at 1 false positive per scan.

The submissions of the UA-2.5D and the submission of ASU-Mayo have been trained with more data than the 20 scans provided by the challenge as training test. They are among the best performing methods, suggesting the need for more training data. This is the rationale for the organizers of the challenge to open of 51 more annotated CT scans, totaling 91 annotated cases.

Five submissions use traditional machine learning approaches, such as SVMs or adaboost with multiple-instance learning (Mevis, UPM, SVM-Features, LMNIIT, FUM-Mvlab), while three other ones use deep learning (UA-2D, UA-2.5D, ASU-Mayo). Deep learning approaches outperform traditional ones, specially at low false positives per scan (bellow two). Out of the three deep learning approaches, only the UA-2D has been trained using only the 20 cases of the challenge training dataset. Such method still clearly outperforms traditional methods trained with such 20 cases.

Two radically different deep learning approaches have been proposed – a global segmentation network by UA (in both the 2D and 2.5D versions) and the use of planar image-reformats from candidates of ASU-Mayo. Surprisingly, the U-Net outperforms the custom-tailored alternative. The reason might be that the global segmentation network skips the selection of candidate points, thus likely reducing the number of false negatives.

Two systems have been evaluated on other databases, demonstrating significantly better performance on them. ASU-Mayo's algorithm achieved a sensitivity of 83% at 2 false positives per scan in [40], while it achieved only 45.5% at 3.2 fps at this challenge. There are clear differences between the datasets. In [40], there are 326 emboli in 121 CTPAs (2,69 emboli/scan), while in the presented dataset there is a total of 90 emboli in 20 CTPA scans (4,5 emboli/scan). Whether this is due to subject selection or criteria when detecting the emboli is subject of debate. We should emphasize that the 20 test cases used for this work have been analyzed by three independent radiologists and their detections have been adjudicated. Mevis' algorithm achieves a sensitivity of 92% in segmental emboli and 90% in subsegmental emboli at 4.8 false positives / scan in [43], while its sensitivity decrease to 40.0% at 1.35 fps in this database. This might be due to the fact that we include central emboli in this database, while its prior evaluation did not. Further, the decrease of sensitivity and fps is coherent with being operating at another point of the ROC curve.

One limitation of the proposed database and competition is that all the cases selected are positive for PE, making impossible to evaluate the negative predictive value of the algorithms. We are looking into manners of extending the database to include such negative cases. While open datasets of CT scans exist, especially in the context of lung cancer screening, to our knowledge, open datasets of CTPA studies with negative cases are non-existent. The second limitation is the reduced number of scans. We have thus extended the size of the dataset that we are making public. We believe that curating open databases that incorporate new cases is a way to further explore new methodologies and its increased performance based on the findings of our work. We are committed to extend the database with new cases based on the other training datasets curated by the participants on the challenge.

Detailed results of the participations are provided in the supplementary material. The dataset can be downloaded from www.cad-pe.org and from IEEE Dataport at http://dx.doi.org/10.21227/9bw7-6823 .

## VI. Conclusion

Through this work we have generated an open database and clear evaluation criteria for the development of CAD for PE algorithms. Seven algorithms have been generated with this database, and one commercial solution evaluated. Several novel deep learning approaches proposed here have outperformed state of the art in CAD for PE, the best one achieving a per-embolus sensitivity of 75% at 2 false positives per scan. Surprisingly, deep learning methods that operate on the whole image outperform custom-developed ones that input emboli information. Such finding, together with the open database and the evaluation metric can lead further developments of CAD for PE to make it useful in clinical practice.

## Acknowledgments

We would like to thank the IEEE ISBI committee for helping us organizing the CAD-PE challenge. This project has been financially supported by the Comunidad de Madrid, Spain through the M+Vision Consortium. This study was supported by the Spanish Ministry of Science and Innovation (project RTI2018-098682-B-I00). The work of Tajbakhsh, Gotway, and Liang on computer-aided diagnosis of pulmonary embolism has been supported by an ASU-Mayo Clinic seed grant and an NIH grant (R01 HL128785). RSJE was partially funded by the R01-HL116931 grant.